\documentstyle[11pt,newpasp,twoside,epsf]{article}
\def\edcomment#1{\iffalse\marginpar{\raggedright\sl#1\/}\else\relax\fi}
\reversemarginpar

\begin{document}
\title{Watching the Birth of Super Star Clusters}
 \author{Jean L. Turner}
\affil{Division of Astronomy and Astrophysics, UCLA, Los Angeles, CA 90095-1562 USA}
\author{Sara C. Beck}
\affil{Department of Physics and Astronomy, Tel Aviv University, Ramat Aviv, Israel}

\begin{abstract}
Subarcsecond infrared and radio observations yield important information 
about the formation of super star clusters from their surrounding gas. We discuss
the general properties of ionized and molecular gas near young, forming SSCs,
as illustrated by the prototypical young, forming 
super star cluster nebula in the dwarf galaxy NGC~5253. This super star cluster appears
to have a gravitationally bound nebula, and the lack of molecular gas suggests a 
very high star formation efficiency, consistent with the formation of a large, bound
cluster.
\end{abstract}

\section{Introduction}
While the first suggestions that young analogues to globular clusters
exist in the local universe came from ground-based imaging 
(Arp \& Sandage 1985), the age of the super star cluster 
really dawned with the subarcsecond 
imaging capabilities of HST, which allows super star clusters (SSCs) to be resolved in nearby
galaxies (e.g., O'Connell et al. 1995, Whitmore \& Schweizer 1995) 
The Antennae galaxies alone have thousands of
young (Myr) SSCs (Whitmore et al. 1999.) Some of these clusters are 
revealed only by infrared imaging (Vigroux et al. 1996); presumably these are the 
youngest regions.  Clues to the mystery of how large and potentially bound
clusters form lie in the environments of the
youngest, embedded regions, which are visible only in the radio and infrared.

That large extragalactic compact HII regions exist and can be detected was first inferred 
for the starburst in M83 by Turner, Ho, \& Beck (1987)
on the basis of a high Brackett line/radio continuum ratio, 
although their 8\arcsec\ (160 pc at M83) infrared beam
did not allow them to define the nature of the unexpected Brackett excess (cf. Thompson 1987). 
With arcsecond mid-infrared imaging, Telesco and Gezari (1992) isolated a compact 12 
$\mu$m source that they suggested was a forming globular cluster in M82. Pina et al.
(1992) and Keto et al. (1993)  
found a similar strong and compact mid-infrared source in the starburst galaxy, NGC~253.

High spatial resolution IR imaging and spectroscopy with
HST and large ground-based telescopes and the expanded high
 frequency imaging capability at the VLA now allow the study 
of nebulae in the infrared and radio at subarcsecond resolution, corresponding 
to scales of $\sim$1 pc in local galaxies. 
With this resolution one can now study the star formation processes
in individual regions.
Due to time and space limitations we will focus on the impact of these observations on
 issues in the formation of large bound clusters as
 exemplified by the well-studied  young embedded SSC in the dwarf galaxy NGC 5253.

\section{Radio Continuum Imaging of the Supernebula in NGC 5253}

Improvements in sensitivity and phase calibration at high frequencies at the VLA
 now allow the detection of  young, compact HII regions in nearby galaxies. 
Free-free emission, with its flat spectrum, is most easily detected at wavelengths
shorter than 2~cm, where it dominates over synchrotron emission, which
declines with frequency. The other advantage of high 
frequency observing is that the resolutions
are higher at these frequencies; beams as small as 50 mas are now possible
with the VLA using the Pie Town link. We can now image parsec-scale structures 
within nearby starbursts. 

NGC 5253 is a peculiar dwarf elliptical galaxy (Caldwell \& Phillips 1989),
with a starburst that has created numerous star clusters within the past 2.5 to 50 Myr 
(Calzetti et al. 1997, Tremonti et al. 2001.) There is abundant and extended H$\alpha$
emission (Walsh \& Roy 1987). Comparison of H$\alpha$ and radio continuum (Beck et al.
1996) indicate that extinctions are high toward the radio peak (Calzetti et al. 1997). 
At a distance of only 3.8 Mpc, we are able to easily resolve the youngest
parts of the starburst in NGC 5253.

\begin{figure}
%\vskip 2.5in
\plotone{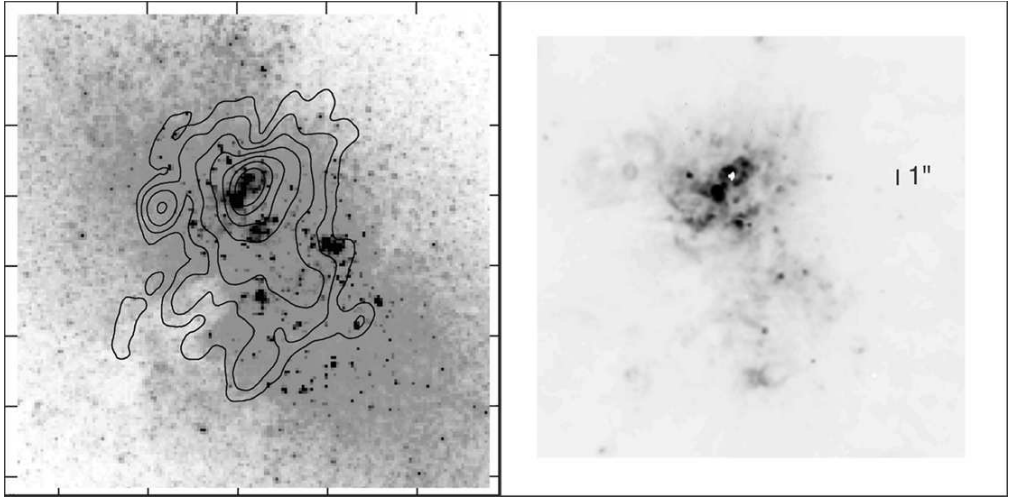}
\caption{Radio continuum source in NGC 5253. (Left) Low resolution
6~cm image (Turner et al. 1998) atop HST I band image. (Right) VLA 2 cm 
image at 0.2\arcsec\ resolution (in white: Turner et al. 2000), on HST H$\alpha$ image.
HST images from Calzetti et al. 1997.}
\end{figure}

Subarcsecond VLA imaging of NGC 5253 at 2 and 1.3 cm revealed a tiny but
 luminous nebula (Figure 1).
Our most recent Pie Town image has a resolution of 50 mas and shows that the nebula
is slightly elliptical, 0.7 pc by 0.9 pc in diameter, and optically thick at 2 cm,
 with a turnover frequency ($\tau$=1) of 15 GHz. This tiny nebula contains at 
least 4000 O stars, and possibly as many as a million stars total, with a 
luminosity of $1\times 10^9~\rm L_\odot$.  About half of this luminosity 
emerges in mid-infrared dust continuum emission (Gorjian, Turner, \& Beck 2001). 
The high density implied
by its high turnover frequency suggests that this ``supernebula" is very young. We 
cannot now say how young, although comparison with Galactic compact HII
 regions would indicate that it is less than 1 Myr old.

\section{What's Going on in There? Can We Know? A Gravity-Bound Nebula}

In order to study the kinematics of the embedded  ``supernebula" in NGC 5253, spectroscopy 
in the infrared or radio is required. Radio recombination lines were detected by Mohan, Ananthamariah, \& Goss (2001), confirming the basic radio continuum findings of Beck
et al. (1996) and Turner et al. (1998).
Brackett recombination lines at 2.17 and 4.05 $\mu$m were detected using NIRSPEC at 
Keck (Turner et al. 2003). These lines also confirm the radio continuum results. The
Brackett line emission in this galaxy is overwhelmingly dominated by the supernebula.

The Brackett lines reveal another interesting and unique property of the supernebula.
The Brackett linewidths are $\rm 75~km\,s^{-1}$, FWHM, consistent with the radio 
recombination lines of Mohan et al. (2001). This is remarkably narrow for the size and
mass of this cluster. For a nebula with diameter 0.7 pc, the escape velocity is 
$\rm 40~km\,s^{-1}$ for a cluster of 4000 O stars alone; for a Salpeter IMF down to
M stars, the escape velocity could be as high as $\rm 120~km\,s^{-1}$, with virial
linewidths only slightly smaller. The supernebula is actually in or close to gravitational 
equilibrium. Nebulae in the Galaxy are not confined by gravity, nor is the more
evolved and diffuse nebula 30 Doradus. This HII region has very different---probably not expanding and possibly static---dynamics. 
%And there are obscured HII regions with similar dynamics in other starburst galaxies. 

It is worthwhile to pause and reflect on the meaning of this small linewidth in the presence
of 4000 O stars within a volume only 1 pc across.  
It is difficult enough to form a single, solitary O star (see the contribution
by H. W. Yorke in this volume), given their violent tendencies,  propensity toward outflow (Figure 2a, Shepherd \& Kurtz 1999) and general windiness.  Outflows, which start in 
the accretion phase, probably continue
throughout a goodly fraction of an O star's main sequence lifetime (Yorke \& Sonnhalter 
2002), and radiation-driven winds certainly do. 
If the O star is massive enough, it may go through a luminous blue variable phase,
such as observed in the star $\eta$ Car, and could lose as much as $10^{-3}~\rm
M_\odot~ yr^{-1}$ (Figure 2b, Morse et al. 1998). Next is the Wolf-Rayet phase,
which occurs for massive stars of a certain age (2-3 Myr); Wolf-Rayet ``ring" nebulae
have sizes of a few pc and velocities up to $100 ~\rm km\,s^{-1}$ (Figure 2c, Gruendl
et al. 2000). The ultimate manifestation of the nastiness of an O star is the 
expulsion of its outer layers in a supernova explosion (Figure 2d, from Hughes et al.
2000). What riotous events are transpiring within the little supernebula? Given
the confinement of gas, can we tell? Confinement of gas in itself could have 
implications for the dynamical evolution of the cluster, and could facilitate 
stellar collisions and perhaps even stellar mergers (Bonnell \& Bate 2002.)

\begin{figure}
\plotone{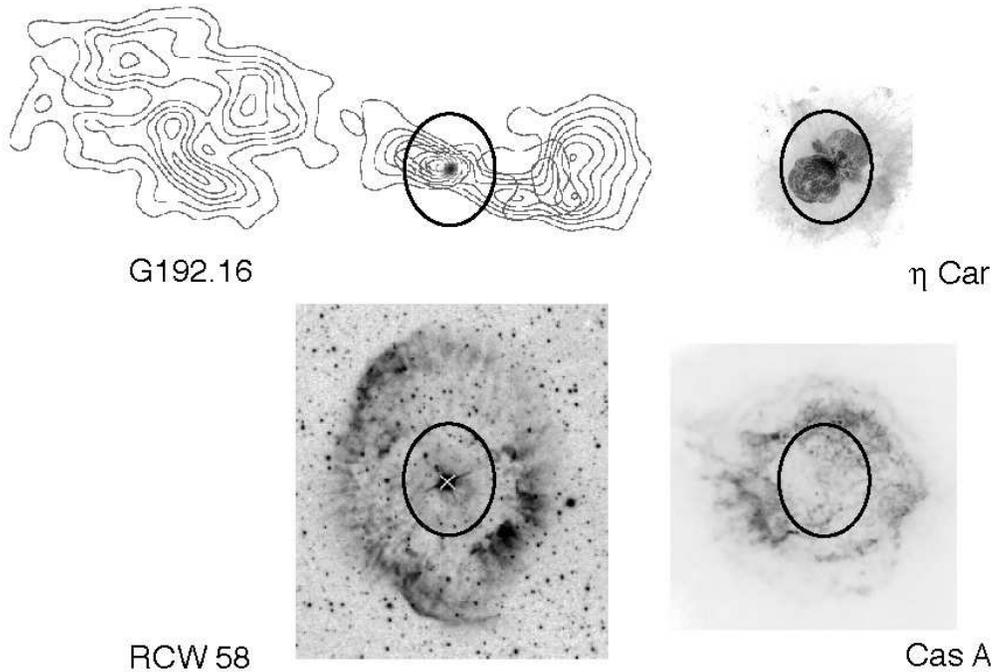}
\caption{The many different ways that O stars can be nasty, from birth to death. The 
ellipse marks the approximate size of the supernebula, with a mean radius
of 0.7 pc. Top left:
the CO outflow in G192.16-3.82, age less than 1 Myr, adapted from Shepherd \& Kurtz 1999. Top right:
the luminous blue variable $\eta$ Car, estimated age 1 Myr, from Morse et al. 1996. Lower
left:  Wolf-Rayet nebula, estimated age 2-3 Myr, adapted from Gruendl \& Chu 2000. 
Lower right: Cas A
supernova remnant,  estimated age (from ZAMS) 5-10 Myr, from Hughes et al. 1999.}
\end{figure}

\section{Formation of a Bound Cluster Must be Efficient, and Is}

To form a bound star cluster, at least 50\% of the original gas mass must end up in stars. Star 
formation efficiencies in the Galaxy are $\sim$1-2\%  on the $\sim 50$ pc sizescales of GMCs 
(Lada, Margulis, \& Dearborn 1984), although efficiencies on sizescales comparable to the 
cluster size are higher.  This is consistent with the lack of young bound clusters in the Galaxy. 

In NGC 5253 we observe a stream of molecular gas falling into the galaxy (Figure 3:
Meier, Turner, \& Beck 2002). There is a similar infalling stream in Henize 2-10 
(Kobulnicky et al. 1995). Most of the
molecular gas is separated spatially and kinematically from the supernebula and from the optical star clusters near it; the infalling gas cannot be associated
causally with the current forming star cluster. There is a small amount of gas, $\sim 4 \times 10^5
~\rm M_\odot,$ within the galaxy proper (Cloud D). 
This can be compared to an estimated $0.7-1.2\times10^6~\rm
M_\odot$ in stars. Star formation has nominally been $\sim$ 60--75\% efficient in NGC~5253.
There are sizeable uncertainties in the gas mass, which is based on two different CO lines,
due to the unknown metallicity dependence of the Galactic conversion factor, and to the
stellar IMF (the lower efficiency corresponds to a lower mass cutoff of 1 $\rm M_\odot$).
However, the gas mass estimated from dust emission is in agreement with the CO mass, 
to within  30\% (Meier et al. 2003, in prep.), so we expect the high efficiency is real. 

\section{Summary}

High resolution radio and infrared observations in nearby galaxies have found luminous and compact nebulae excited by obscured super star clusters.  The IR emission of these nebulae can be a significant, or even dominant, part of the total galactic flux.  In NGC 5253,  there are so many young stars in such a small volume that the gravitational attraction appears to be slowing or stopping the rapid expansion of the nebula. The star formation efficiency also appears to be
quite high in this galaxy, $\sim 60-75$\%, which is consistent with the formation of bound
clusters.  Star formation in SSC-starburst galaxies does not much resemble that closer to home. 

\begin{figure}
\plotone{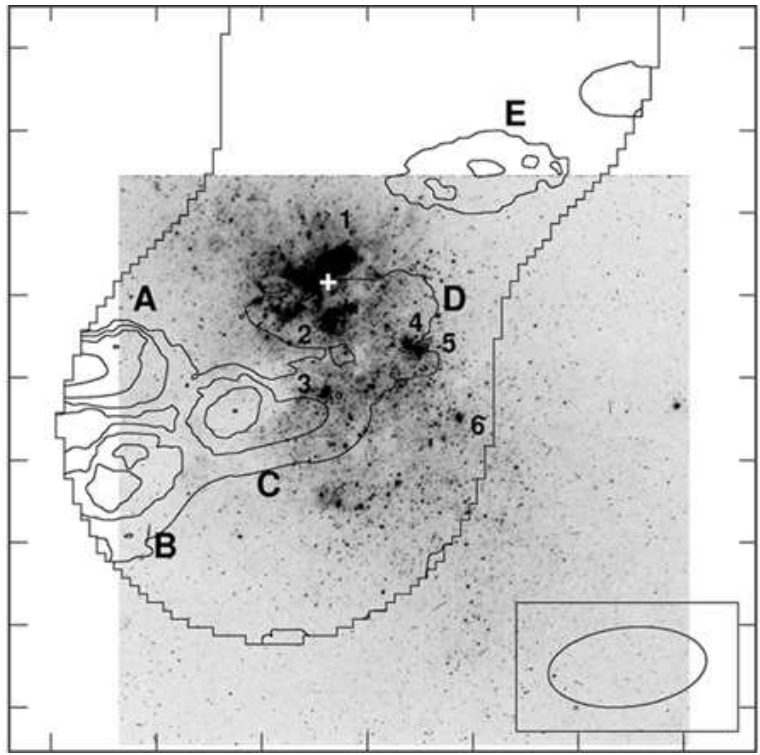}
\caption{CO in NGC 5253, from Meier et al. 2002. The white cross marks the supernebula. There are 5 clouds of gas and dust in this figure, marked A through E. Only
cloud D is resident within the galaxy, while the other clouds are falling in. Cloud D has
a mass of $\sim 4 \times 10^6~ \rm M_\odot.$}
\end{figure}

\acknowledgments
The authors would like to acknowledge the contributions of P.T. P. Ho, D. S. Meier,
J. H. Lacy, V. Gorjian, L. Crosthwaite, D. S. Meier, J. Larkin, \& I. McLean  to this work. 
This research is supported by NSF Grant 0307950.


\begin{references}

\reference Beck, S. C., Turner, J. L., Ho, P. T. P., Kelly, D., \& Lacy, J. H. 1996, \apj, 457, 610
\reference Bonnell, I. A., \& Bate, M. R. 2002, \mnras, 336, 659
\reference Caldwell, N., \& Phillips,  M. M. 1989, \apj, 338, 789
\reference Gruendl, R. A., Chu, Y.-H., Dunne, B. C., \& Points, S. D. 2000, \aj, 120, 2670
\reference Hughes, J. P., Rakowski, C. E., Burrows, D. N., \& Slane, P. O. 2000, \apj, 528, 
L109
\reference Kobulnicky, H. A., Dickey, J. M., Sargent, A. I., Hogg, D. E., \& Conti, P. S. 
2000, \aj, 120, 1273
\reference Lada, C. J., Margulis, M., \& Dearborn, D. 1984, \apj, 285, 141
\reference Meier, D. S., Turner, J. L., \& Beck, S. C. 2002, \aj, 124, 877
\reference Morse, J., Davidson, K., Bally, J., Ebbets, D., Balick, B., \& Frank, A. 1998, \aj,
116, 2443
\reference O'Connell, R. W., Gallagher, J. S. III, Hunter, D. A., \& Colley, W. N. 1995, \apj, 
446, L1
\reference Pina, R. K., Jones, B., Puetter, R. C., \& Stein, W. A. 1992, \apj, 401, L75
\reference Shepherd, D. S., Kurtz, S. E. 1999, \apj, 523, 690
\reference Telesco, C. M., \& Gezari, D. Y. 1992, \apj, 395, 461
\reference Thompson, R. I. 1987, \apj, 321, 153
\reference Turner, J. L., Beck, S. C., Crosthwaite, L. P., Larkin, J. E., McLean, I. S., \&
Meier, D.S. 2003, Nature, 423, 621
\reference Turner, J. L., Beck, S. C., \& Ho, P. T. P. 2000, \apj, 532, L109
\reference Turner, J. L., Ho, P. T. P., \& Beck, S. C. 1987, \apj, 313, 644
\reference Turner, J. L., Ho, P. T. P., \& Beck, S. C. 1998, \aj, 116, 1212
\reference Vigroux, L., et al. 1996, \aap, 315, L93
\reference Walsh, J. R. \& Roy, J.-R. 1987, \mnras, 239, 297
\reference Whitmore, B. C., \& Schweizer, F. 1995, \aj, 109, 960
\reference Whitmore, B. C., Zhang, Q., Leitherer, C., Fall, S. M., Schweizer, F., \& Miller, B. W. 
1999, \aj, 118, 1551
\reference Yorke, H. W., \& Sonnhalter, C. 2002, \apj, 569, 846

\end{references}
\end{document}